\documentclass[
aps,prl,showpacs,superscriptaddress,numerical,amsmath,amssymb,floatfix,reprint]{revtex4-2}
\usepackage[T1]{fontenc}
\usepackage[utf8]{inputenc}
\usepackage[
pdffitwindow=true,
colorlinks=true,
frenchlinks=false,
linkcolor=blue,
anchorcolor=blue,
citecolor=blue,
filecolor=blue,
urlcolor=blue,
bookmarks=true,
bookmarksopen=true,
bookmarksnumbered=true,
bookmarksopenlevel=1,
plainpages=false,
pdfpagelayout=TwoPageLeft,
pdfpagelabels=true,
breaklinks]{hyperref}
\usepackage[main=english]{babel}
\usepackage[dvipsnames]{xcolor}
\usepackage[per-mode=symbol,separate-uncertainty]{siunitx}
\usepackage{graphicx}
\usepackage{dcolumn}
\usepackage{color}
\usepackage{graphicx,wrapfig,lipsum}
\usepackage{bm}

\usepackage{color}
\usepackage{ulem}
\usepackage{float}
\usepackage{chemformula}

\begin{document}
	\title{Chiral phonons and phononic birefringence in ferromagnetic metal - bulk acoustic resonator hybrids 
	}
	\author{M.~M\"uller}
	\email{manuel.mueller@wmi.badw.de}
	\affiliation{Walther-Mei{\ss}ner-Institut, Bayerische Akademie der Wissenschaften, 85748 Garching, Germany}
	\affiliation{Technical University of Munich, TUM School of Natural Sciences, Physics Department, 85748 Garching, Germany}
	
	\author{J.~Weber}
	\affiliation{Walther-Mei{\ss}ner-Institut, Bayerische Akademie der Wissenschaften, 85748 Garching, Germany}
	\affiliation{Technical University of Munich, TUM School of Natural Sciences, Physics Department, 85748 Garching, Germany}
	
	\author{F.~Engelhardt}
	\affiliation{Max Planck Institute for the Science of Light, 91058 Erlangen, Germany}
	\affiliation{Department of Physics, University Erlangen-Nuremberg, 91058 Erlangen, Germany}
	\affiliation{Institute for Theoretical Solid State Physics, RWTH Aachen University, 52074 Aachen, Germany}
 \author{V.A.S.V.~Bittencourt}
 \affiliation{ISIS (UMR 7006), Universit\'e de Strasbourg, 67000 Strasbourg, France }
	\author{T.~Luschmann}
	\affiliation{Walther-Mei{\ss}ner-Institut, Bayerische Akademie der Wissenschaften, 85748 Garching, Germany}
	\affiliation{Technical University of Munich, TUM School of Natural Sciences, Physics Department, 85748 Garching, Germany}
	\affiliation{Munich Center for Quantum Science and Technology (MCQST), 80799 Munich, Germany}
	\author{M.~Cherkasskii}
 \affiliation{Institute for Theoretical Solid State Physics, RWTH Aachen University, 52074 Aachen, Germany}

	\author{S.T.B.~Goennenwein}
	\affiliation{Department of Physics, University of Konstanz, 78457 Konstanz, Germany}
	\author{S.~Viola Kusminskiy}
	\affiliation{Institute for Theoretical Solid State Physics, RWTH Aachen University, 52074 Aachen, Germany}
	\affiliation{Max Planck Institute for the Science of Light, 91058 Erlangen, Germany}
	
	\author{S.~Geprägs}
	\affiliation{Walther-Mei{\ss}ner-Institut, Bayerische Akademie der Wissenschaften, 85748 Garching, Germany}

	\author{R.~Gross}
	\affiliation{Walther-Mei{\ss}ner-Institut, Bayerische Akademie der Wissenschaften, 85748 Garching, Germany}
	\affiliation{Technical University of Munich, TUM School of Natural Sciences, Physics Department, 85748 Garching, Germany}
	\affiliation{Munich Center for Quantum Science and Technology (MCQST), 80799 Munich, Germany}
	
	\author{M.~Althammer}
	\email{matthias.althammer@wmi.badw.de}
	\affiliation{Walther-Mei{\ss}ner-Institut, Bayerische Akademie der Wissenschaften, 85748 Garching, Germany}
	\affiliation{Technical University of Munich, TUM School of Natural Sciences, Physics Department, 85748 Garching, Germany}
	\author{H.~Huebl}
	\email{hans.huebl@wmi.badw.de}
	\affiliation{Walther-Mei{\ss}ner-Institut, Bayerische Akademie der Wissenschaften, 85748 Garching, Germany}
	\affiliation{Technical University of Munich, TUM School of Natural Sciences, Physics Department, 85748 Garching, Germany}
	\affiliation{Munich Center for Quantum Science and Technology (MCQST), 80799 Munich, Germany}
	\date{\today}
		\begin{abstract}
Magnomechanical devices, in which magnetic excitations couple to mechanical vibrations, have been discussed as efficient and broadband microwave signal transducers in the classical and quantum limit. We experimentally investigate the magnetoelastic coupling between the ferromagnetic resonance (FMR) modes in a metallic \ch{Co25Fe75} thin film, featuring ultra-low magnetic damping as well as sizable magnetostriction, and standing transverse elastic phonon modes in sapphire, silicon and gadolinium gallium garnet by performing broadband FMR spectroscopy at cryogenic temperatures. For all these substrate materials, we observe an interaction between the resonant acoustic and magnetic modes, which can be tailored by the propagation direction of the acoustic mode with respect to the crystallographic axes. We identify these phonon modes as transverse shear waves propagating with slightly different velocities with relative magnitudes of $\Delta v/v\simeq10^{-5}$, i.e., all substrates show phononic birefringence. Upon appropriately choosing the phononic mode, the hybrid magnomechanical system enters the Purcell enhanced coupling regime.

		\end{abstract}
		\maketitle
		\textit{Introduction ---} Phonons, the quantized excitations of elastic waves in solids, are a crucial concept for a vast variety of solid-state phenomena. Recently, they came into focus in the field of quantum science due to potential applications in quantum memories\cite{Chu2017, Chu2018, Hann2019,Wallucks2020}, quantum transducers\cite{Han2022,McKenna2020, Engelhardt2022a,Shen2020a,Jiang2020, Arnold2020}, and quantum sensors\cite{Bienfait2020, Jain2022, Delsing2019,Satzinger2018, Potts2021}. The main focus then is on standing elastic waves in mechanical resonators with high quality factors\cite{Aspelmeyer2014, OConnell2010} realized e.g. in the form of high harmonic overtone bulk (BAW)\cite{Goryachev2012,Goryachev2018} or surface acoustic wave (SAW) resonators\cite{Chu2017, Chu2018, Andersson2022, Ekstrom2017}. These resonators support both longitudinal and transversal (shear) elastic standing waves, typically with linear polarization. However, superpositions of phonons based on shear waves can also carry angular momentum. This enables the transfer or storage of angular momentum as well as the conversion between circularly polarized and linearly polarized shear waves, allowing for the realization of phononic spin valves\cite{An2020,An2022}. Experiments investigating these concepts require the excitation and detection of phonons with a defined circular polarization. To this end, the intrinsic chirality of a magnetic resonance mode such as the Kittel mode in combination with the magnetoelastic interaction represents a promising approach for detecting and driving such phonons\cite{Flebus2017, Sharma2022a,Bittencourt2022, Wachter2021}. Magnetic materials combined with BAW resonators, which provide long-lived excitations at high overtone frequencies, are therefore an ideal testbed to study the excitation of phonons with angular momentum \cite{An2020, An2022,Schlitz2022,Streib2018,Sato2021,Hatanaka2022, Peria2021,Keshtgar2014}. While in early experiments \cite{An2020, An2022, Schlitz2022} the ultra-low magnetization damping material yttrium iron garnet (\ch{Y3Fe5O12}, YIG) grown lattice matched on gadolinium gallium garnet (\ch{Gd3Ga5O12}, GGG) substrates is used, the following aspects remained unaddressed: (i) What requirements must the phonon dispersion relations of a given crystalline symmetry fulfill to allow for the transport of angular momentum or the conversion of chiral to linear polarized phonons and vice versa? (ii) Can the phonon modes involved be experimentally resolved? (iii) Is the excitation concept unique for epitaxial YIG films on GGG substrates or can it be generalized to other material systems, e.g., polycrystalline metallic thin films? 
		
		Here, we address all of these questions and show that polycrystalline metallic magnetic thin films are well suited for the excitation of high overtone BAWs in sapphire (\ch{Al2O3}), GGG and silicon (Si). This provides clear evidence that the excitation scheme is generic. We extract this information from the resonant interaction of the ferromagnetic excitations with the standing elastic waves from the BAWs. We find that at least two propagation velocities or phonon modes must be considered in our substrates suggesting that the phonon propagation direction with respect to crystallographic axes is of key importance to carry angular momentum and to convert chiral to linearly polarized phonons. This important birefringence effect establishes novel concepts for the transformation and control of chiral phonon modes.

\begin{figure}[tbh]	
		\centering
		\includegraphics[width=1.0\columnwidth, clip]{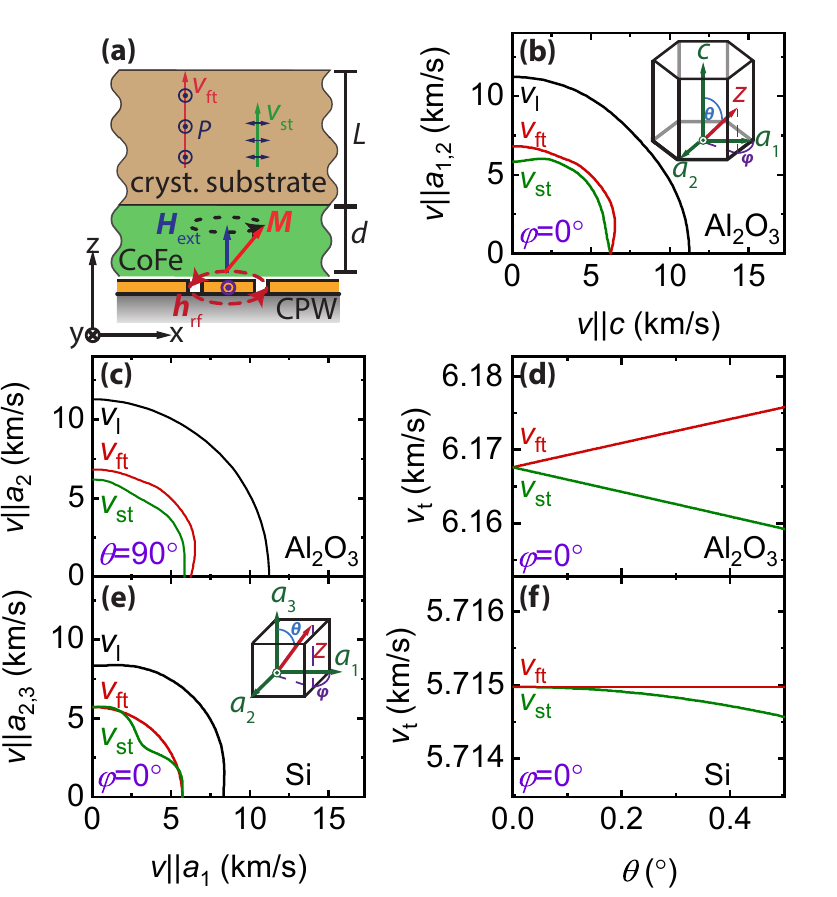}
		\caption{(a)\,Schematic of the sample composed of a metallic CoFe thin film on a crystalline substrate mounted on a coplanar waveguide. The transverse acoustic waves can split into fast and slow modes with velocities $v_{\mathrm{ft}}$ and $v_{\mathrm{st}}$, respectively (red and green arrows), and different polarization direction $\boldsymbol{P}$ (blue arrows).\,(b)\,Group velocities of the longitudinal and transverse acoustic phonons of hexagonal \ch{Al2O3} in the $\boldsymbol{a_{1,2}}$-$\boldsymbol{c}$-plane. The length of the vector from the origin to the colored lines gives the magnitude of the velocities and its direction the propagation direction in the $\boldsymbol{a_{1}}$-$\boldsymbol{c}$-plane. The vector $\boldsymbol{\mathrm{z}}$ defines the phonon propagation direction with $\theta$ and $\varphi$ representing the azimuthal and polar angles of $\boldsymbol{\mathrm{z}}$ with respect to the hexagonal
lattice vectors $\boldsymbol{\mathrm{c}}$ and $\boldsymbol{\mathrm{a_1}}$.\,(c)\,Phonon group velocities of \ch{Al2O3} in the $\boldsymbol{a_{1}}$-$\boldsymbol{a_{2}}$-plane for $\theta=90^\circ$.\,(d) Calculated splitting of the slow and fast transverse mode velocities $v_{\mathrm{st}}$ and $v_{\mathrm{ft}}$ in \ch{Al2O3} as function of $ \theta$ ($\varphi=0^\circ$).\,(e)\,Group velocities of the longitudinal and transverse acoustic phonons of cubic Si in the $\boldsymbol{a_{1}}$-$\boldsymbol{a_{2,3}}$-plane. The vector $\boldsymbol{\mathrm{z}}$ defines the phonon propagation direction with respect to the cubic lattice vectors $\boldsymbol{a_{\mathrm{i}}}$. \,(f)\,Calculated splitting of $v_{\mathrm{st}}$ and $v_{\mathrm{ft}}$ for $\varphi=0^\circ$ as a function of $\theta$.}
		\label{Fig: 1}
		\end{figure}

		\textit{Qualitative discussion ---}
		We consider ferromagnetic metallic (FM) \ch{Co25Fe75} (CoFe) thin films deposited on crystalline substrates (see Fig.\,\ref{Fig: 1}\,(a)). When subjected to a sufficiently large magnetic field $H_{\mathrm{ext}}$ oriented along the film normal ($\boldsymbol{\mathrm{z}}$-direction), the magnetization of CoFe aligns parallel to $H_{\mathrm{ext}}$. Using a microwave drive, we excite the ferromagnetic resonance (FMR) Kittel mode, whose frequency depends on $H_{\mathrm{ext}}$. Due to the finite magnetoelastic coupling (MEC)\cite{Weiler2011, Weiler2012,An2020,An2022,Schlitz2022}, the magnetization dynamics generates a high frequency stress field and hence elastic modes with the same frequency and helicity as the magnonic mode. As the magnetic thin film is elastically coupled to the substrate, the elastic modes (phonons) can also propagate therein. This process is called phonon pumping\cite{Streib2018}. For the geometry chosen in our experiment (see Fig.\,\ref{Fig: 1}(a)), the excited phonons are exclusively the transverse acoustic phonons as the projection of the magnetization vector $\boldsymbol{M}$ on the field direction is constant in the linear regime and hence no longitudinal acoustic phonons are generated along the out-of-plane direction $\boldsymbol{\mathrm{z}}$\,\cite{Streib2018, Sato2021}. Due to our experimental setting with a thin magnetic film acting as transducer on top of a substrate, the properties of the standing waves are dominated by the phonon dispersion relation of the substrate material, which depends on the details of the substrate's crystal structure\cite{ashcroft2011solid,GrossMarx+2022, Wolfe1998}.
		
Owing to the excitation scheme just described, these phonons have wave numbers close to the center of the Brillouin zone, where the dispersion relation is to a good approximation linear, and, hence, group and phase velocities are identical. However, depending on the symmetry of the crystal, these velocities are usually anisotropic, i.e., they depend on the propagation direction ($\boldsymbol{\mathrm{z}}$-direction in our setting) relative to the crystallographic axis (cf. Fig.\,\ref{Fig: 1}(b)), as described by the Christoffel equation\cite{Fedorov1968,Jaeken2016}. Figures\,\ref{Fig: 1}(b) and (c) show the calculated phonon propagation velocities solving the Christoffel equation\cite{Jaeken2016} for the exemplary case of hexagonal \ch{Al2O3}\cite{Tefft1966}, where the anisotropy of the velocities reflect the symmetry of the underlying crystal structure. For shear waves with a propagation direction $\boldsymbol{\mathrm{z}}$ along the hexagonal $\boldsymbol{c}$-direction, the velocities of fast and slow transverse phonon modes, $v_{\mathrm{ft}}$ and $v_{\mathrm{st}}$, are identical. However, if $\boldsymbol{\mathrm{z}}$ is not parallel to the $\boldsymbol{c}$-axis, this degeneracy is lifted even for small angles $\theta$ between $\boldsymbol{\mathrm{z}}$ and $\boldsymbol{c}$ as displayed in Fig.\,\ref{Fig: 1}(d). Moreover, for small $\theta$-values, the $\varphi$-dependence of $v_{\mathrm{ft}}/v_{\mathrm{st}}$ can be neglected. This lifting of degeneracy for $v_{\mathrm{ft}}$ and $v_{\mathrm{st}}$ is not unique to hexagonal crystals. Figures\,\ref{Fig: 1}(e) and (f) show the corresponding phonon velocity anisotropy for Si (diamond structure)\cite{Malica2020} and the small angle $\theta$ dependence, respectively.
		
		As depicted in Fig.\,\ref{Fig: 1}\,(a), the propagating phonons meet reflective boundary conditions imposed by the finite thickness $L+d\simeq L$ of the sample composed of substrate ($L$) and magnetic thin film ($d$). This leads to the formation of standing waves. For $L\simeq 510\,\mu$m this corresponds to a mode spacing in the MHz range for transverse acoustic modes propagating with velocities $\simeq 6$\,km/s. As we excite the FMR in the CoFe layer at GHz frequencies, we operate this BAW resfonator in the high-overtone regime. Here, the resonance frequency of an elastic standing wave mode with mode number ${n}$ can be well approximated by\cite{Sato2021}
		\begin{equation}
		 f_{n,i}=n/[2(d/\tilde{v}_{\mathrm{t}}+L/v_{i})]. \label{eq:modefrequencies}
		\end{equation} 
		Here, $i=\mathrm{st},\,\mathrm{ft}$ and $\tilde{v}_{\mathrm{t}}$ denotes the transverse phonon velocity of the magnetic thin film. Whenever the frequency of the FMR is resonant with one of the standing wave modes of the BAW, we can excite the elastic mode via MEC. Therefore, we expect a change of the FMR signature due to phonon pumping\cite{Streib2018, Sato2021} at $f_n$ and thus the FMR absorption line to show a frequency-periodic modification, where the periodicity is given by the free spectral range $f_{\mathrm{FSR}}=f_{n+1}-f_n=f_1$ \cite{An2020,An2022,Schlitz2022}. Note that the FMR absorption line is sensitive to all elastic modes, which can be excited with the stress field created via MEC in the FM film. Realistic substrates will have a small but finite miscut angle $\theta_\mathrm{m}$, which denotes the angle between the surface normal $\boldsymbol{\mathrm{z}}$ and the respective crystallographic axis. Therefore, we expect the observation of non-degenerate propagation velocities $v_{\mathrm{ft}}$ and $v_{\mathrm{st}}$ along $\boldsymbol{\mathrm{z}}$, which manifest themselves as a superposition of frequency periodic intersections of the FMR with periods given by the respective free spectral ranges $f_{\mathrm{1,ft}}$ and $f_{\mathrm{1,st}}$, respectively, if the linewidths of the phononic modes are sufficiently narrow to resolve them. This allows one to determine the resonance spectra of the BAW resonator and, hence, according to Eq.\,\eqref{eq:modefrequencies} the group velocities of the excited acoustic phonon modes. Thus, this technique represents a technologically simple, but sensitive tool for the investigation of these resonators.		
		
		\textit{The experiment ---} We deposit a Pt(3\,nm)/Cu(3\,nm)/CoFe(30\,nm)/Cu(3\,nm)/Ta(3\,nm) multilayer stack via dc magnetron sputtering on a $510$\,$\mu$m thick (0001)-oriented \ch{Al2O3} substrate, a $675$\,$\mu$m thick (001)-oriented high-resistivity Si substrate, and a $380$\,$\mu$m (001)-oriented GGG substrate, which are all polished on both sides. The seed layer of the CoFe film composed of a Pt(3\,nm)/Cu(3\,nm)-bilayer is thin enough to maintain good elastic coupling between the substrates and the CoFe layer, but is required to ensure optimal magnetization damping properties of CoFe\cite{Edwards2019, Flacke2019, Schoen2016}. The top Cu(3\,nm)/Ta(3\,nm) layers prevent the CoFe thin film from oxidation. To analyze the MEC, we perform broadband FMR experiments in a magnet cryostat\cite{Maier-Flaig2018}. The sample is mounted face-down onto a coplanar waveguide and we record the frequency-dependent complex microwave transmission parameter $S_{21}$ as a function of the out-of-plane magnetic field $H_{\mathrm{ext}}$ using a vector network analyzer.
		\begin{figure}[tbh]	
			\centering
			\includegraphics[width=1.0\columnwidth, clip]{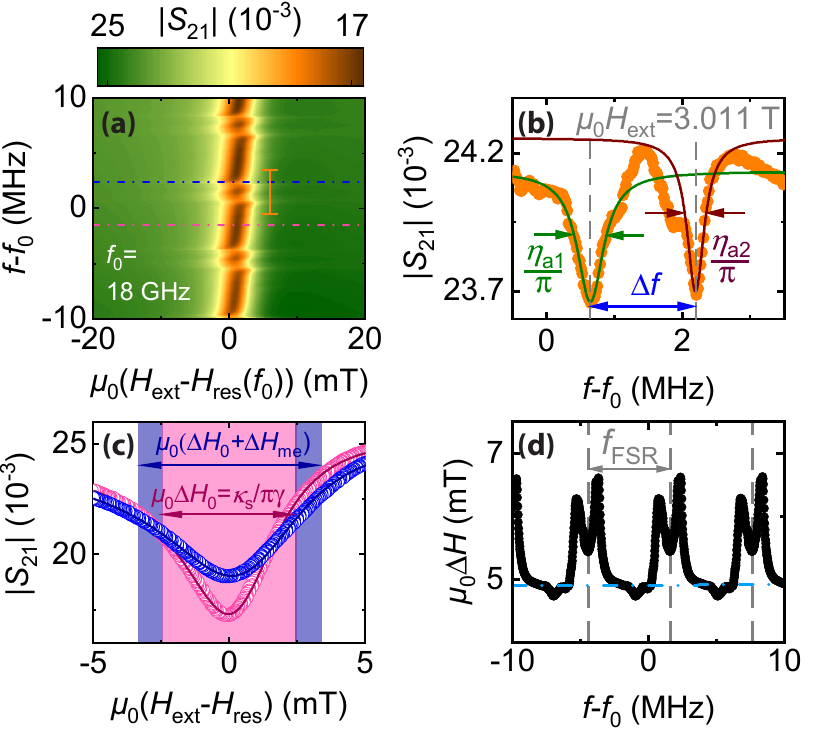}
			\caption{(a)\,Microwave transmission magnitude $|S_{21}|$ as a function of $f-f_0$ around $f_0=18$\,GHz and $H_{\mathrm{ext}}$ around $\mu_0 H_\text{res}(f_0) =3.005$\,T recorded at $T=5\,$K. Panel (b) shows the $|S_{21}|$ along the orange vertical line in (a). We choose $\mu_0H_{\mathrm{ext}}=3.011$\,T to analyze the unperturbed linewidth of the elastic modes and obtain the BAW resonator decay rates $\eta_{\mathrm{a}1,2}$ by fitting two lorentzian lines to the data (green and brown lines). The blue arrow indicates the spacing of the double peaks $\Delta f$. Panel (c) shows $|S_{21}|$ for the blue (magenta) horizontal dashed lines in (a), which correspond to the resonant (off-resonant) case of the $n^{\mathrm{th}}$ bulk elastic resonance interacting with the FMR. On resonance, we observe a linewidth broadening $\Delta H_{\mathrm{me}}$ due to the MEC. In panel (d) $\Delta H(f)$ is plotted as a function of $f$. The gray dashed lines mark the free spectral range $f_{\mathrm{FSR}}$ and the light blue dashed line represents the linear dispersion of $\Delta H$ with an offset $H_{\mathrm{inh}}$.}
			\label{Fig: 2}
		\end{figure}
		Fig.\,\ref{Fig: 2}(a) shows $|S_{21}|$ as a function of $H_{\mathrm{ext}}$ and microwave frequency $f$ around $f_0=18$\,GHz recorded at $T=5$\,K for CoFe deposited on an \ch{Al2O3} substrate. We observe the characteristic FMR of the CoFe layer, which shows a distinct, frequency periodic pattern, which we interpret this as the signature of the interaction between the magnon mode and the high overtone BAW mediated by MEC and elastic coupling at the interface to the substrate (cf.\,Refs.\,\cite{An2020, Schlitz2022}). The observed double peak structure with a frequency periodicity of $f_{\mathrm{FSR}}\approx6.04$\,MHz is in good agreement with the expected $f_{\mathrm{FSR}}\approx 6.05$\,MHz from Eq.\,(\ref{eq:modefrequencies}) in \ch{Al2O3} using the material parameters of CoFe and \ch{Al2O3} ($d=30$\,nm, $L=510\,\mu$m, $v_{\mathrm{ft,st}}\approx6.17$\,km/s\cite{Tefft1966} and $\tilde{v}_{\mathrm{t}}=3.17$\,km/s\cite{Schwienbacher2019}). The detection of two neighboring phonon resonances with a frequency separation of $\Delta f\approx1.40$\,MHz suggests the presence of two non-degenerate transverse acoustic phonon velocities in the substrate differing by $\Delta v_{\mathrm{t}}\simeq$0.5\,m/s. We can resolve this small difference in the propagation velocities due to the narrow linewidth of the acoustic modes at low temperatures. Following Ref.\,\cite{Schlitz2022}, we extract the undisturbed linewidth of the elastic modes from a linescan at constant $\mu_0H_{\mathrm{ext}}=3.011$\,T detuned from the FMR (orange vertical line in Fig.\,\ref{Fig: 2}(a)) and the FMR linewidth from a fixed frequency linescan detuned from the acoustic modes (dashed magenta line in Fig.\,\ref{Fig: 2}(a)). The respective data are presented in Fig.\,\ref{Fig: 2}(b) and (c). We find the BAW decay rates $\eta_{\mathrm{a1,2}}/(2\pi)\approx(0.23\pm0.02)\,\mathrm{MHz},\,(0.16\pm0.01)$\,MHz and the FMR damping rate $\kappa_{\mathrm{s}}/(2\pi)\approx (69.0\pm0.1)$\,MHz from the half-width at half maximum linewidth of the elastic resonances and FMR, respectively. From the BAW decay rates, we calculate an BAW decay length of $2\pi\delta_{1,2}=v_{\mathrm{t}}/\eta_{\mathrm{a}1,2}\approx2.6\,\mathrm{cm},\,3.8\,$cm$> 2L$ at $f_0=18$\,GHz which allows for the formation of standing waves in the \ch{Al2O3} substrate. To quantify the MEC, we calculate the coupling rate $g_{\mathrm{MEC}}$ with the model in Ref.\,\cite{An2020} (see SM\cite{Supplements}). We obtain the coupling rate $g_\mathrm{MEC1,2}/2\pi\simeq6.0$\,MHz at $f_0=18\,$GHz assuming a magnetoelastic constant of $B=15.7\times 10^6\,\mathrm{J/m^3}$, a mass density of $\rho=8110\,\mathrm{kg/m^3}$ and a transverse velocity of $\tilde{v}_{\mathrm{t}}=3.17\,$km/s for CoFe\cite{Schwienbacher2019}. 
		In this regime, we can understand additional losses of the FMR as an effectively Purcell enhanced damping\cite{Herskind2009, Huebl2013}, which we can quantify by studying the magnetic field linewidth of the FMR when resonant with an acoustic mode as a function of $f$ as presented in panel (d). Using this representation, we can clearly identify the resonance frequency of the BAW modes in our sample. Obviously, the study of $\Delta H(f)$ provides a very sensitive probe to detect phonon modes in our experiments ($\Delta v_{\mathrm{t}}/v_{\mathrm{t}}=\eta_{\mathrm{a}}/(2\pi f_0)\simeq10^{-5}$ comparable to Brillouin spectroscopy\cite{Minami2006}). In addition, this analysis also allows in principle to investigate the intrinsic magnetization damping mechanisms in CoFe. The light blue dashed line represents the linear increase of the FMR-linewidth $\Delta H$ due to Gilbert damping with damping parameter $\alpha=(2.8\pm0.1)\cdot10^{-3}$ . The offset results from inhomogeneous broadening characterized by $H_{\mathrm{inh}}=(1.6\pm0.3)$\,mT. Both $\alpha$ and $H_{\mathrm{inh}}$ are extracted from broadband FMR experiments over a frequency range from $5$\,GHz to $50$\,GHz (see SM\cite{Supplements}).

		\begin{figure}[tbh]	
			\centering
			\includegraphics[width=1.0\columnwidth, clip]{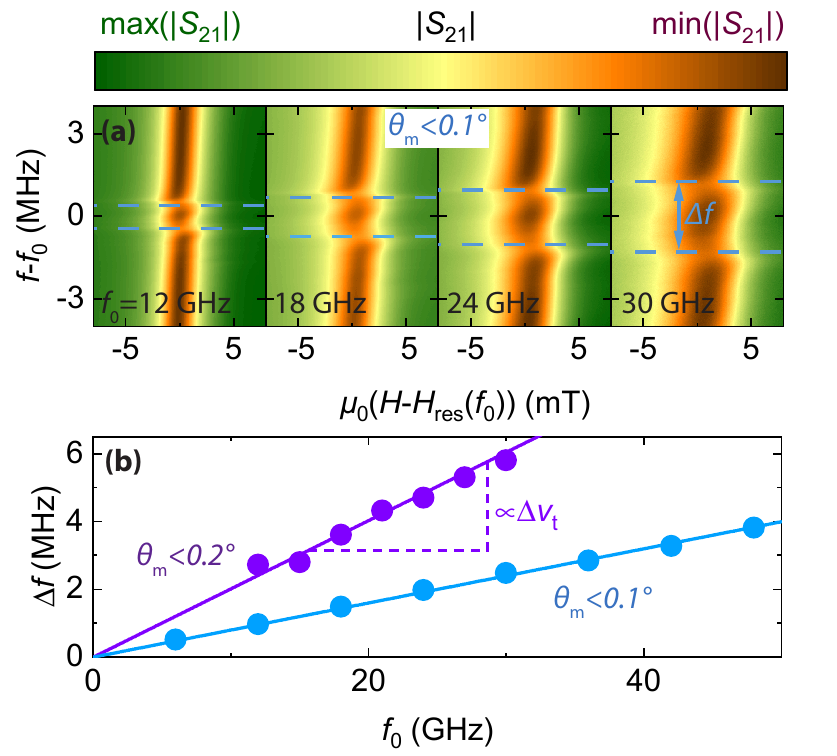}
			\caption{(a)\,Microwave transmission magnitude $|S_{21}|$ as a function of frequency $f$ and $H_{\mathrm{ext}}$ recorded in narrow intervals around $f_0$ and $H_{\mathrm{res}}(f_0)$ at $T=5$\,K for a CoFe thin film deposited on a $c$-axis \ch{Al2O3} substrate with a miscut $ \theta_{\mathrm{m}}<0.1^\circ$. The values of $H_{\mathrm{res}}(f_0)$ as well as $\mathrm{min}(|S_{21}|)$ and $\mathrm{max}(|S_{21}|)$ at the respective $f_0$ are given in Tab.\,SI in the SM\cite{Supplements}. Note that the frequency spacing of the panels in (a) is 6\,GHz. Panel (b) displays the observed frequency splitting $\Delta f$ for this sample as well as for a second sample of CoFe on a $c$-axis \ch{Al2O3} substrate with a miscut of $ \theta_{\mathrm{m}}<0.2^\circ$ as function of $f_0$ (blue and purple circles) together with linear fits to extract the velocity difference of the two transverse modes $\Delta v_{\mathrm{t}}$. }
			\label{Fig: 3}
		\end{figure}
		
	 Next, we investigate the splitting of the acoustic modes on a broader frequency scale by studying two modes associated with two different group velocities for driving frequencies $f_0$ ranging between 12 and 30 GHz (see Fig.\,\ref{Fig: 3} and Tab.\,SI in the SM\cite{Supplements}). We find that the frequency separation $\Delta f$ of the BAW resonances from the two phonon velocities scales linearly with $f_0$. This behavior is expected from Eq.\,(\ref{eq:modefrequencies}) and should also manifest itself as two different $f_{\mathrm{FSR}}$ values for both modes. However, the precision of the latter is not sufficient to determine the velocity difference of $\Delta v_\mathrm{t}\approx0.5$\,m/s observed here. For sapphire, $v_{\mathrm{ft}}$ is equal to $ v_{\mathrm{st}}$ for sound propagation along the $\boldsymbol{c}$-axis (cf. Fig.\,\ref{Fig: 1}(b)). Assuming that the presence of two slightly different propagation velocities $v_{\mathrm{ft}}$ and $v_{\mathrm{st}}$ originates from a slight misalignment $\theta_{\mathrm{m}}$ between the propagation direction $\boldsymbol{\mathrm{z}}$ and the $\boldsymbol{c}$-axis of the sapphire crystal, we can estimate $\theta\approx0.017^\circ$ (see Fig.\,\ref{Fig: 1}\,(d)). Here, we neglect the $\varphi$-dependence, which is valid for small $\theta$. This $\theta$ is compatible with the suppliers miscut specifications of the sapphire substrate of $\theta_\mathrm{m} <0.1^\circ$. To test this conjecture, we performed experiments on a second CoFe thin-film on a different sapphire substrate with $\theta_\mathrm{m} < 0.2^\circ$ (see purple circles in Fig.\,\ref{Fig: 3}\,(b) and Fig. S2 in the SM\cite{Supplements}), where we find a slightly larger scaling of $\Delta f$ with $f_0$ indicating a higher $\Delta v_\mathrm{t}\approx1.1\,$m/s corresponding to $\theta=0.037^\circ$.

\begin{figure}[tbh]	
			\centering
			\includegraphics[scale=1.0, clip]{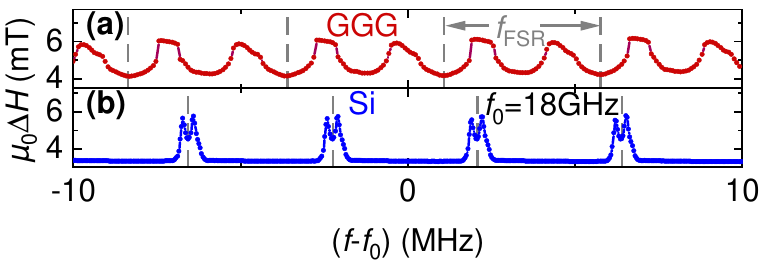}
			\caption{Fitted FMR linewidth $\Delta H(f)$ recorded at $T=5$\,K as function of $f$ around $f_0=$18\,GHz for CoFe deposited on various substrate materials. Pronounced MEC peaks are observed for CoFe deposited on (a) a $380\,\mu$m thick GGG (111) substrate ($\theta_\mathrm{m}<1.0^\circ$), and (b) a $675\,\mu$m thick Si (100) substrate ($\theta_\mathrm{m}<0.5^\circ$). Gray dashed lines indicate the $f_{\mathrm{FSR}}$.}
			\label{Fig: 4}
		\end{figure}

In a last step, we investigate whether these observations are unique to CoFe on hexagonal \ch{Al2O3}. To this end, we deposit CoFe films on cubic GGG and Si substrates. Fig.\,\ref{Fig: 4} shows the fitted FMR linewidth $\Delta H(f)$ as a function of $f$.
	We find that the double peak features in $\Delta H(f)$ are not unique to hexagonal \ch{Al2O3} substrates (see Fig.\,\ref{Fig: 2}(d)) but are also observed for CoFe grown on (a) $380\,\mu$m thick (111)-oriented cubic GGG and (b) on $675\,\mu$m thick (100) oriented cubic Si. As expected, the splitting $\Delta f$ and the free spectral range $f_\mathrm{FSR}$ of the overtone resonances differ for the various substrates, as the absolute values of $v_\mathrm{st}$ and $v_\mathrm{ft}$ are different for the various substrates (see SM\cite{Supplements}). In addition, the magnitude of both the off- and on-resonant FMR linewidth is comparable for all substrates. This suggests that (i) the magnetization damping and thus the magnetic properties of the CoFe film are consistent for the various substrates, (ii) that the acoustic damping as well as the thicknesses of the substrates are similar, and (iii) that the underlying excitation mechanism is governed by the material parameters of CoFe \cite{Schwienbacher2019, Sato2021}.

	\textit{Conclusion ---} 
	We report the magnetoelastic coupling between the ferromagnetic resonance mode in a ferromagnetic metal and the transverse acoustic phonon modes of high-overtone bulk acoustic resonators in the Purcell enhanced regime. Using thin polycrystalline ferromagnetic CoFe films to drive the acoustic excitations, we explore \ch{Al2O3}, Si, and GGG as phononic host materials. As the chirality of the ferromagnetic resonance modes can excite circularly polarized elastic shear waves in the substrates which carry angular momentum, this excitation scheme is considered ideal for the investigation of phononic angular momentum transport. The two transverse shear wave velocities can also differ depending on the crystallographic symmetry and the phonon propagation direction with respect to the crystal axes. For \ch{Al2O3} and Si substrates we find velocity differences in the m/s range and thus these substrates can be considered as birefringent for phonons \cite{Born1999}. The combination of the simple and efficient ferromagnetic excitation schemes with substrates hosting long-lived phonons thus offers an unique opportunity to explore angular momentum transport by phonons and the controlled conversion between linear and circular polarized phonons via birefringence\cite{Psarobas2014}. Finally, this method should allow for the determination of the helicity of phonons even in the quantum limit.
	
	\nocite{nye1985physical, Yamamoto1941, Klokholm1982}

\section{Acknowledgments}
We acknowledge financial support by the Deutsche
Forschungsgemeinschaft (DFG, German Research Foundation) via Germany’s Excellence Strategy EXC-2111-390814868 and SFB 1432 (Project-ID 425217212). F. E., V.A.S.V. B, M.C. and S.V.K thank Sanchar Sharma for useful discussions and acknowledge funding from the Bundesministerium für Bildung und Forschung (BMBF , Grant No. 16KIS1590K) and from the Deutsche Forschungsgemeinschaft (DFG, German Research Foundation) through Project- ID 429529648–TRR 306 QuCoLiMa (“Quantum Cooperativity of Light and Matter”).	

	\bibliography{library}

\end{document}